\documentstyle[pre,aps,amsmath,epsf,tighten,floats]{revtex}    

\begin{document}
\draft
\title{Order statistics of the trapping problem }

\author{ Santos B. Yuste and Luis Acedo }
\address{Departamento de F\'{\i}sica, Universidad  de  Extremadura,
E-06071 Badajoz, Spain}

\date{\today}

\maketitle
\begin{abstract}
When a large
number $N$ of independent diffusing particles are placed upon a site of a  $d$-dimensional
Euclidean lattice randomly occupied by a concentration $c$ of  traps, what is the $m$th moment
$\langle t^m_{j,N} \rangle$ of the time $t_{j,N}$ elapsed until the first $j$ are trapped?
An exact answer is given in terms of the probability $\Phi_M(t)$
that no particle of an initial set of $M=N, N-1,\ldots, N-j$ particles is trapped by
time $t$. The Rosenstock approximation is used to evaluate $\Phi_M(t)$,
and it is found that for a large range of trap concentracions  the $m$th
moment of $ t_{j,N}$ goes as $x^{-m}$ and its variance as $x^{-2}$, $x$ being $\ln^{2/d}
(1-c) \ln N$.  A rigorous asymptotic expression (dominant and two corrective
terms) is given for $\langle t^m_{j,N}\rangle$ for the one-dimensional lattice.
\end{abstract}

\section{Introduction}

Statistical problems related to the diffusion of a single random walker in
a medium with traps have been subject of intense research during the last decades
\cite{Hughes,Weiss,TrapLibVarios,TrappingN1,DV,DV-BBB,Zumofen,Blumen}.
Usually it is assumed  that the statistical
properties of this single ($N=1$) random walker are representative of the statistical ensemble.
However, there are multiparticle ($N>1$)  problems that
can not be analyzed in terms of the single walker theory.  An example is
the  number $S_N(t)$ of distinct
sites visited  (or territory explored)  up to time $t$ by $N$ independent random walkers all
starting from the same origin
\cite{LarraldeN,LarraldeP,OtrosSNt,SNtEucRapid,SNtEucRegular,SNtFrac}.
Another multiparticle problem of interest, that as we will see is closely related to
that of the territory explored,  is the description of the order statistic of the
diffusion processes, i.e., the estimate of the time at which the $j$th particle of an
initial set of $N$ particles all starting from the same origin is trapped.

The
order-statistic problem when the traps are arranged on a (hyper) sphere  (i.e., an
absorbing boundary at a fixed distance) has been thoroughly studied
\cite{JSP-Lindenberg-Weiss,OS-Yuste,DK,YAK-submitted}. In this paper we
consider the more difficult  problem in which the traps are arranged randomly
(``the trapping problem'') in a $d$-dimensional Euclidean medium. A related
problem, in which the $N$ particles are placed on the left of a one-sided random
distribution of traps on a one-dimensional lattice, has been investigated in
Ref.\ \cite{OneSided}. It is interesting to note that recent advances in optical
spectroscopy \cite{SingMol} make it possible to monitor this kind of multiparticle dynamic
process. Indeed, the simultaneous tracking of $N\gg 1$ fluorescently labeled
particles and the analysis of the diffusive motion of the particles {\em individually}
is a useful recent technique for characterizing heterogeneous microenvironments (in
particular for samples dynamically changing in time such as biological samples)
\cite{MultiTracking}.
 An useful feature of the order statistics
approach is that it allows one to infer properties of the diffusive system (diffusion
constant, number of diffusing particles, concentration of traps,  effective dimension of
the diffusive substrate,\ldots) from only the analysis of the behavior of those
particles that are first trapped.   This could be an advantage when it is impractical or impossible to wait until all the reaction is over .

The order statistics of the trapping process will be described by means of the
probability   $\Phi_{j,N}$  that $j$ particles  of the initial set of $N$ diffusing
particles  have been trapped, and the other $N-j$
have survived, by time $t$. In this paper we consider that all the
particles start from the same origin which is free of traps.
The moments of the time $t_{j,N}$  at which the $j$th
particle of the initial set of $N$ particles  is trapped will be calculated from
$\Phi_{j,N}$ .   This probability $\Phi_{j,N}$ will be given in terms
of the survival probability $\Phi_M(t)\equiv \Phi_{0,M}(t)$ that no particle of an
initial set of $M$ ($M=N,N-1,\ldots,N-j$) has been absorbed by time $t$. This last
quantity will be estimated by means of the Rosenstock approximation using expressions
for $S_N(t)$ calculated  in Refs.\ \cite{SNtEucRapid,SNtEucRegular,SNtFrac}.
It should be noted that our approach to the order statistics of the diffusion process in 
the presence of randomly placed traps is different from that used
\cite{JSP-Lindenberg-Weiss,OS-Yuste,DK,YAK-submitted} for a fixed configuration of
traps.  What makes the two problems completely different, and hence the way of
solving them, is that  for the case with a given configuration of traps the
probability that $N$ particles are trapped by time $t$ is simply the $N$th power of
the probability for a single particle. This simplifying result does not hold when many configurations of randomly placed traps are considered.

The plan of the paper is as follows.
In Sec.\ \ref{sec:OSTP}  we  deduce the main formulas that describe the order statistics of the
trapping process: we relate $\Phi_{j,N}$  to $\Phi_N(t)$  and  $\langle t^m_{j,N}\rangle$
to $\Phi_{j,N}(t)$.
In Sec.\ \ref{sec:SNtmom} we show that the ratio between the variance of
$S_N(t)$ and   $\langle S_N(t)\rangle^2$  goes roughly as $(\ln N)^{-2}$ for large $N$.
This suggests that the Rosenstock approximation for $\Phi_N(t)$ can lead to good results even when $N$ is large. This is checked in Sec.\ \ref{sec:OSRA}, where we also obtain asymptotic expressions (the main term) for $\langle t^m_{j,N}\rangle  $ and the variance.
The procedure, based on the Rosenstock approximation, does not provide analytic asymptotic
corrective terms for $d\ge 2$, although we show that numerical integration is feasible leading to excellent results. 
However, in Sec. \ref{sec:Rig1D}, for the one-dimensional lattice  we are able to find  a
rigorous asymptotic expression (up to the second-order corrective term) for 
$\langle t^m_{j,N}\rangle$  for large  $N$.
Some remarks and the conclusions are presented in Sec.\ \ref{sec:Conclu}.

\section{Order statistics of the trapping process}
 \label{sec:OSTP}
Let us first show  how to obtain  $\Phi_{j,N}(t)$  from  $\Phi_M(t)$ with $M=N, N-1, \ldots,
N-j$. Let $\Psi_{j,N}(t)$ be the probability that $j$ random walkers of the initial
set of $N$ have been absorbed by time $t$ by a given configuration of traps and let
$\Psi(t)\equiv\Psi_{0,N}(t)$ be the probability that no single random walker has been absorbed by time $t$ by this configuration of traps. 
 Taking into account that $\tbinom{N}{j}$ is the number of different groups of $j$ particles that can be formed from a set of $N$, one finds
\begin{equation} 
\Psi_{j,N}(t)=
\binom{N}{j} \left(1-\Psi\right)^j  \Psi^{N-j},
\end{equation}
or, using the binomial expansion,
\begin{equation}
\Psi_{j,N}(t)=
\binom{N}{j}\sum_{m=0}^j (-1)^{j-m} \binom{j}{m}  \Psi^{N-j+j-m} .
\label{PsijN}
\end{equation}
Averaging over different configurations of traps and taking into account that
 $\Phi_N(t)= \langle\Psi^N(t)\rangle$  and
 $\Phi_{j,N}(t)= \langle\Psi_{j,N}(t)\rangle$, we get
\begin{equation}
\Phi_{j,N}(t) = (-1)^j
\binom{N}{j}
  \Delta^j \Phi_N(t)  ,
\label{PhijN}
\end{equation}
where  the backward difference
formula for the $j$th derivative
 \begin{equation}
 \Delta^j \Phi_N(t)=\sum_{m=0}^j
(-1)^m
\binom{j}{m}
 \Phi_{N-m}(t)
 \label{DifferenceFormula}
 \end{equation}
 has been used.
 The difference formula in Eq.\ (\ref{PhijN}) can be approximated by the
derivative: \begin{equation}
\Phi_{j,N}(t)  \simeq (-1)^j
\binom{N}{j}
  \frac{d^j}{dN^j} \Phi_N(t)
\label{PhijNaprox}
\end{equation}
when  $j\ll N$. 
Let  $h_{j,N}(t)$ be the probability that the
$j$th absorbed particle of the initial set of $N$ disappears during the time interval $(t,t+dt]$.
This quantity is related to $\Phi_{j,N}(t)$ by
\begin{equation}
h_{j+1,N}(t)
=h_{j,N} -\frac{d}{dt} \Phi_{j,N}(t)
=-\frac{d}{dt}\sum_{m=0}^j \Phi_{m,N}(t)
\label{hj1N}
\end{equation}
with $h_{0,N}=0$.
Then, the $m$th moment of the  time at which the $j$th particle is  trapped is given by
\begin{equation}
\langle t_{j,N}^m  \rangle = \int_0^\infty t^m h_{j,N}(t) dt
\label{tjN}
\end{equation}
or, using Eq.\ (\ref{hj1N}) and integrating by parts,  by
\begin{equation}
\langle t_{j+1,N}^m  \rangle = \langle t_{j,N}^m \rangle+
m \int_0^\infty t^{m-1} \Phi_{j,N}(t) dt
\label{tjNa}
\end{equation}
with
\begin{equation}
\langle t_{1,N}^m \rangle = m \int_0^\infty t^{m-1} \Phi_{N}(t) dt          .
\label{t1Na}
\end{equation}
Using Eq.\ (\ref{PhijN}), Eq.\ (\ref{tjNa}) becomes
\begin{equation}
\langle t_{j+1,N}^m \rangle =
\langle t_{j,N}^m \rangle +
(-1)^j
\binom{N}{j}
 \Delta^j \langle t_{1,N}^m \rangle         .
\label{tjNb}
\end{equation}
Thus,  the order statistics of the trapping problem is described from $\langle t_{1,N}^m \rangle $ only.
However,   when $N$ and $j$ are large,  Eq.\ (\ref{tjNb}) is hardly useful numerically  
because the quantities $\langle t_{1,N-r}^m \rangle$, that are added and subtracted (and almost cancelled) to obtain
the $j$th difference derivative $\Delta^j \langle t_{1,N}^m \rangle$ have to be calculated, then, with extraordinary accuracy (which is not easy; see Secs.\ \ref{sec:OSRA} and \ref{sec:Rig1D}) in order to get a reasonable estimate for the small quantity $(-1)^j(\langle t_{j+1,N}^m \rangle -\langle t_{j,N}^m \rangle)$ from the multiplication of the tiny quantity $\Delta^j \langle t_{1,N}^m \rangle$ by the huge binomial coefficient.
In Sec.\ \ref{sec:OSRA} we will show how one can surmount,
at least partially,  this difficulty.

\section{Moments of the number of distinct sites visited by $N$ random walkers}
\label{sec:SNtmom}

 The main purpose of this section is to show that for large $N$ one can approximate
    $\langle S^2_N(t) \rangle$ by $\langle S_N(t) \rangle^2$. In other words, we will show that
the ratio     $\text{Var}(S_N)/\langle S_N \rangle^2$ is small for large $N$ and that it decreases
when $N$ increases. In fact, we will show that the simulation results are compatible  with the
conjecture made in \cite{OneSided} that    $\left[\text{Var}(S_N)\right]^{1/2}/\langle S_N \rangle \sim 1/\ln N$.
 These results make it very plausible that the Rosenstock
approximation of order zero is a reliable method for estimating the survival
probability $\Phi_N(t)$ for not too long times and small concentrations.   
This will be analyzed in Sec.\ \ref{sec:OSRA}.

The problem of evaluating  $\langle S^m_N(t) \rangle$  for $N=1$  has been intensively studied
since it was posed by Dvoretzky and Erd\"os  \cite{Hughes,Weiss,DE}.  In 1992,  Larralde
{\em et al.} \cite{LarraldeN,LarraldeP} addressed the problem for $N\gg 1$ and $m=1$ on Euclidean media. 
They disclosed the existence of three time regimes:
a very short-time regime  [$t \ll t_{\times}\sim \ln( N)/\ln( d)$], or regime I, in which there are so
many particles at every site that all the nearest neighbors of the already visited sites are
reached at the next step, so that the number of distinct sites visited grows as the volume of an
hypersphere of radius $t$, $\langle S_N(t) \rangle \sim  t^d$;   a very long-time regime
($t_{\times}^{\prime}\ll t$), or  regime III, that is the  final stage in which the walkers  move
far away from each other so that their trails (almost) never overlap and $\langle S_N(t) \rangle
\sim N \langle S_1(t) \rangle$; and an intermediate   regime   ($ t_{\times}\ll t \ll
t_{\times}^{\prime} $), or regime II, in which there exists a non-negligible probabilitiy 
of the trails of the particles overlapping. Of course,  regime III does not exist for $d=1$ (
$t_{\times}^{\prime}=\infty$).  For $d=2$, $t_{\times}^{\prime}=\exp(N)$, and for $d\geq 3$,
$t_{\times}^{\prime}=N^{2/(d-2)}$. In the simulations carried out in this paper and for the values of $N$  we are interested in ($N\gg 1$), the particles spend  most of the time
inside regime II, and regime III is never reached.

 For regime II  it has been found that
\cite{SNtEucRapid,SNtEucRegular}
\begin{equation}
\label{SNt}
\langle S_N(t) \rangle  \approx \widehat S_N(t) (1-\Delta)
\end{equation}
with
\begin{eqnarray}
\label{SNtgorro}
\widehat S_N(t)&=&v_0 \left(4 D t \ln N \right)^{d/2},  \\
\label{DeltaSN}
\Delta  \equiv  \Delta(N,t) & = &
\frac{1}{2}\sum_{n=1}^{\infty} \ln^{-n} N \sum_{m=0}^n s_m^{(n)} \ln^{m} \ln N  \;
\end{eqnarray}
and where, up to second order ($n=2$),
\begin{eqnarray}
\label{s10}
s_0^{(1)}&=& -d \omega  ,\\
 s_1^{(1)}&=& d \mu   ,\\
 s_0^{(2)}&=&
d\left(1-\frac{d}{2}\right) \left( \frac{\pi^2}{12}+\frac{\omega^2}{2} \right)
-d\left(\frac{d h_1}{2}-\mu \omega\right) ,\\
 s_1^{(2)}&=&  - d \left(1-\frac{d}{2}\right)\mu \omega - d\mu^2 ,\\
s_2^{(2)}&=& \frac{d}{2} \left(1-\frac{d}{2}\right) \mu^2  .
\label{scoef}
\end{eqnarray}
Here $\omega=\gamma+\ln A+ \mu \ln(d/2)$, where $\gamma \simeq 0.577215$ is the Euler
constant, $v_0$ is the volume  of the hyphersphere with unit radius, 
 and  $A$, $\mu$ and $h_1$ are given in Table \ref{table1} for $d=1$,
$2$ and $3$. The diffusion constant is defined by means of the Einstein relation
\begin{equation}
\langle r^2 \rangle \approx 2dDt,
\end{equation}
for large $t$,  with $\langle r^2 \rangle$  being the mean-square displacement of a
single random walker. All the numerical results that appear in this paper are calculated using $D=1/(2d)$. 

\begin{table}
\caption{
 Parameters  that appear  in the asymptotic expression of $S_N(t)$
Eq.\ (\protect\ref{SNt}). The symbol $d$D refers to the
$d$-dimensional simple hypercubic lattice. The parameter $\widetilde p$ is
$\left[ 2(6D\pi)^3/3 \right]^{1/2} p({\bf 0},1)$
\protect\cite{nota1}  where  $p({\bf 0},1)=
\sqrt{6}/(32\pi^3)\Gamma(1/24)\Gamma(5/24)\Gamma(7/24)\Gamma(11/24) \simeq 1.516386$
\protect\cite{Hughes}.
}
\label{table1}
\begin{tabular}{cccc}
Case &  $A$ & $\mu$ & $h_1$  \\
\tableline
1D & $\sqrt{2/\pi}$ & 1/2 & -1 \\
2D & $1/\ln t$		&  1  & -1  \\
3D & $1/(\widetilde p\protect{\sqrt{t}})$ &1&-1/3  \\
\end{tabular}
\end{table}

However, the calculation of higher-order moments of
$ S_N(t)$ poses a problem of completely different order of magnitude that still remains
unsolved.
In Ref. \cite{OneSided}, it was conjectured that  the functional form of $\langle  S^m_N \rangle$
for Euclidean lattices has the same asymptotic structure for all $m$,  namely,  the asymptotic
structure of Eq.\ (\ref{SNt}).
Moreover, it was conjectured that
\begin{equation}
\frac{\text{Var}(S_N)}{\langle  S_N \rangle ^2} \sim \frac{1}{\ln^{2}N}
      \left[1+{\cal O}\left(\frac{\ln^3\ln N}{\ln N}\right)   \right]
\label{conjetura}
\end{equation}
for large $N$, where $\text{Var}(S_N)=\langle  S^2_N\rangle-\langle S_N\rangle^2$ is the
variance of $S_N(t)$. Note that Eq.\ (\ref{conjetura})  implies  $\langle  S^2_N\rangle
= \langle S_N \rangle^2$ up to the first-order asymptotic corrective term, as well as
$\text{Var}(S_N) \sim t^d(\ln N)^{d-2}$ for large $N$.

\begin{figure}
\begin{center}
\leavevmode
\epsfxsize = 7.5cm
\epsffile{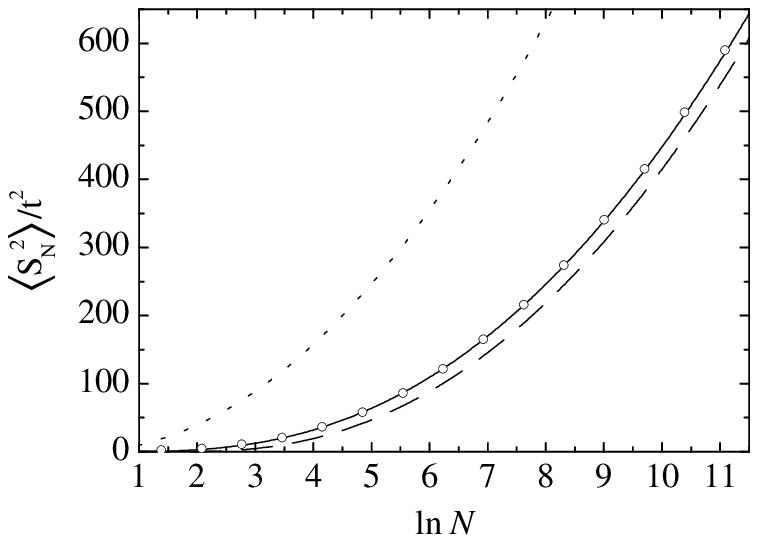}
\end{center}
\caption{
 $\langle S^2_N \rangle/t^2$  versus $\ln N$ for the two-dimensional lattice when $t=400$. The circles are simulation results averaged over $10^5$  configurations for $N=2^2,\ldots,2^{12}$ and
 over $10^4$ configurations for $N=2^{13},\ldots,2^{16}$.
The lines represent $\langle
S_N(t) \rangle^2$ when the main term (dotted line),  first-order approximation (dashed
line) and second-order approximation (solid line) for   $\langle
S_N(t) \rangle$ are used}
\label{fig:2momD2}
\end{figure}

\begin{figure}
\begin{center}
\leavevmode
\epsfxsize = 6cm
\epsffile{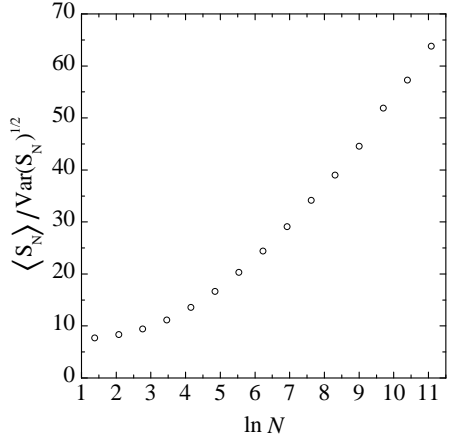}
\end{center}
\caption{
Simulation results for the ratio $\langle S_N  \rangle /\left[\text{Var}(S_N)\right]^{1/2}$ for the two-dimensional
lattice with  $N=2^2, 2^3,\ldots,2^{16}$ and $t=400$. The configurations employed were the same as in 
Fig.\ \ref{fig:2momD2}.
 } \label{fig:Var2D}
\end{figure}

Simulation data  for
$\langle S^2_N(t)\rangle$  for the two-dimensional lattice are compared in Fig.\
\ref{fig:2momD2} with  results obtained from the approximation $\langle
S^2_N(t)\rangle \simeq \langle S_N(t)\rangle ^2$ in which the zeroth-, first- and
second-order asymptotic approximation for $\langle S_N(t)\rangle$ given by Eq.\
(\ref{SNt}) is used.
The large difference between the performance of the
three asymptotic approximations is quite noticeable as well as the excellent result obtained with the second-order approximation.  Similar results (not shown) are found for $d=1$ and $d=3$.
Figure \ref{fig:Var2D} shows  simulation data for the ratio $\text{Var}(S_N)/\langle
S_N \rangle^2$  for the two-dimensional lattice. We see that for large $N$ this ratio
decays roughly as predicted by Eq.\ (\ref{conjetura}).

\section{Order statistics of the trapping process by means of the Rosenstock approximation}
\label{sec:OSRA}

The extended Rosenstock approximation (or truncated cumulant expansion)
first proposed by Zumofen and Blumen \cite{Zumofen}  is a well-known procedure \cite{Hughes,Weiss,TrapLibVarios}
for solving the Rosenstock trapping problem for a single  particle ($N=1$).
Its generalization for estimating the (survival) probabilty $\Phi_N(t)$ that no particle of the initial set of $N$ diffusing
particles has been trapped by time $t$ is straightforward  (details can be
found in Ref.\ \cite{OneSided}) and  we will only quote here those results that are
useful for our objectives.

The zeroth-order Rosenstock approximation  for estimating $\Phi_N(t)$ is given by
\begin{equation}
\label{Ros:zero}
\Phi_N^{(0)}(t)=e^{-\lambda \left\langle S_N(t) \right\rangle  }
\end{equation}
where $\lambda \equiv -\ln (1-c)$ and $c$ is the concentration of traps.
We will write  $\Phi_N^{(0n)}(t)$ to indicate that the $n$th-order approximation for
$\langle S_N(t)\rangle$  [see Eq.\ (\ref{SNt})] is used.
The first-order  Rosenstock approximation is:
\begin{equation}
\label{Ros:first}
\Phi_N^{(1)}(t)=\exp\left[\displaystyle \left \langle S_N(t) \right \rangle
\displaystyle\ln p
\left(1+\displaystyle\frac{\lambda}{2}
\frac{\text{Var}(S_N)}{\langle S_N(t) \rangle} \right) \right].
\end{equation}
Then, the error made by using the zeroth-order Rosenstock approximation can be
estimated:
\begin{equation}
\label{Ros:error}
\Phi_N(t)=\Phi_N^{(0)}(t)\left[1+{\cal O}\left( \lambda^2 \text{Var}(S_N)  \right) \right] .
\end{equation}
Thus, the condition $\lambda^2 \text{Var}(S_N) \ll 1$ guarantees the good performance of the
zeroth-order Rosenstock approximation.
We have found in Sec.\ \ref{sec:SNtmom}  that
$\text{Var}(S_N) \sim t^d(\ln N)^{d-2}$ so that the zeroth-order Rosenstock
approximation works well when $\lambda^2 t^d (\ln N)^{d-2} \ll 1$.
This means that the approximation improves slightly  for $d=1$ and worsens
slightly  for $d=3$ when $N$ increases.  
For long times, the Rosentock approximation eventually breaks
down, the Donsker-Varadhan regime settles in,  and the survival probability decays in a
distinct way known in the literature as Donsker-Varadhan behaviour \cite{DV}.

\begin{figure}
\begin{center}
\leavevmode
\epsfxsize = 7.5cm
\epsffile{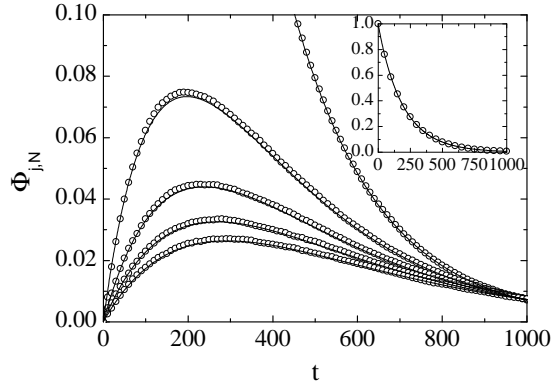}
\end{center}
\caption{
The $j$th survival probability $\Phi_{j,N}(t)$ versus time for (from top to bottom)
$j=0,1,2,3,4$, with $N=1000$ and $c=4\times10^{-4}$ for the two-dimensional lattice.
The lines represent $\Phi_{j,N}^{(02)}(t)$, i.e., the zeroth-order Rosenstock
approximation with $\langle S_N(t)\rangle$ given by the second-order
asymptotic approximation. The circles are simulation results averaged over $10^6$
configurations. Inset: $\Phi_{0,N}(t)$.}
 \label{fig:PhijN}
\end{figure}

Figure  \ref{fig:PhijN} shows  the survival probability $\Phi_{j,N}$ for the
two-dimensional lattice obtained from computer simulations and from Eq.\ (\ref{PhijN}) when the
zeroth-order Rosenstock approximation  $\Phi^{(02)}_{N}(t)$ given by Eq.\
(\ref{Ros:zero}) is used. The agreement is excellent.

Now, we  evaluate $\langle t_{1,N}^m \rangle$ by means of Eq.\ (\ref{t1Na})
approximating the survival probabity $\Phi_N(t)$ by the  the zeroth-order Rosenstock
approximation $\Phi^{(0)}_{N}(t)$  for {\em all} times
: \begin{equation}
\langle t_{1,N}^m \rangle \simeq
m \int_0^\infty t^{m-1} \exp \left[-\lambda \langle S_N(t) \rangle  \right ] dt.
\label{t1Nb}
\end{equation}
Notice that with this approximation we are assuming that, in the integration of Eq.\ (\ref{t1Na})
that leads to  $\langle t_{1,N}^m \rangle$, the relevant contribution comes from the time interval in which the Rosenstock approximation works.
 Next, the expression  for $\langle S_N(t) \rangle $ corresponding to the
 {\em intermediate} time regime is used in Eq.\ (\ref{t1Nb})  for {\em all} times.
This approximation is reasonable  if the integrals of $ m t^{m-1} \Phi_N(t) $   on the
intervals $[0,t_\times]$ and $[t_\times^\prime,\infty]$ are negligible versus $\langle
t_{1,N}^m \rangle$. As $t_\times\sim \ln N$,  the approximation concerning the first
interval is good as long as $(\ln N)^m \ll \langle t_{1,N}^m \rangle$,
i.e. [see Eq.\ (\ref{t1Nc}) below], as long as
 $\lambda \ll (\ln N)^{-d}$.
For $t\ge t_\times^{\prime}$, one has
$ \langle S_N(t)\rangle \sim N \langle S_1(t) \rangle$,
so that the  approximation regarding the interval   $[t_\times^\prime,\infty]$ is good
when  $\lambda \exp(N)\gg 1$ for $d=2$ and $\lambda N^3 \gg 1$ for $d=3$.
Inserting the main asymptotic term of  $\langle S_N(t) \rangle$, namely, $\langle S_N(t)
\rangle \approx v_0 (4Dt\ln N)^{d/2}$, into Eq.\ (\ref{t1Nb}) one gets, after a simple
integration,  a zeroth-order approximation for the $m$th moment of $t_{1,N}$:
 \begin{equation}
 \langle t_{1,N}^m \rangle \simeq
\frac{\Gamma(1+2m/d)}{\left(\lambda v_0 \right)^{2m/d}} \frac{1}{(4D\ln N)^m} .
\label{t1Nc}
\end{equation}
The corrective terms of $\langle S_N(t) \rangle $ are not used in Eq.\ (\ref{t1Nb})
because their time dependence for the two- and three-dimensional cases impedes analytical integration.

\begin{figure}
\begin{center}
\leavevmode
\epsfxsize = 6.5cm
\epsffile{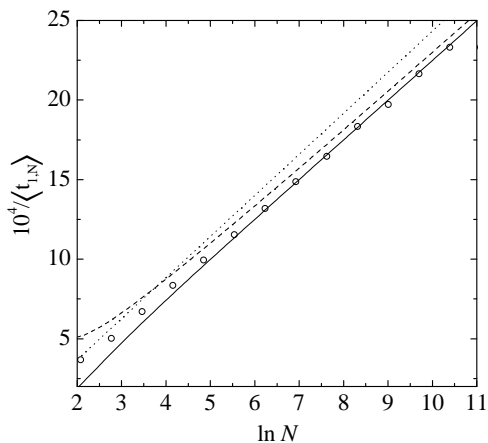}
\end{center}
\caption{ The function $10^4/\langle t_{1,N}\rangle$ versus $\ln N$ for the
one-dimensional lattice  with concentration of traps $c=8\times 10^{-3}$ and
$N=2^3,2^4,\ldots,2^{16}$.  We plot simulation results averaged over $10^5$
configurations(circles) and the asymptotic approximations of order 0 (dotted line),
order 1 (dashed line) and order 2 (solid line).}
\label{fig:t1N1D} \end{figure}

\begin{figure}
\begin{center}
\leavevmode
\epsfxsize = 6.5cm
\epsffile{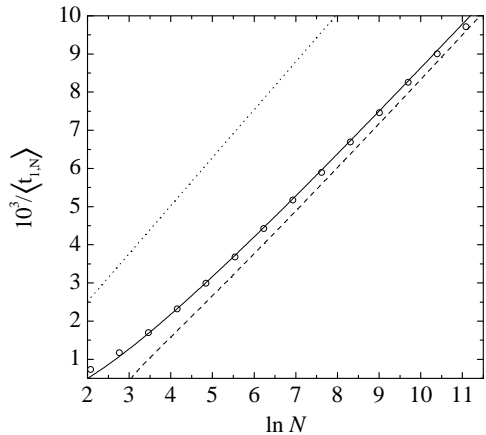}
\end{center}
\caption{ The function $10^3/\langle t_{1,N}\rangle$ versus $\ln N$ for the
two-dimensional lattice  with  $c=4\times 10^{-4}$ and
$N=2^3,2^4,\ldots,2^{16}$. The simulation results are averaged over
$10^5$ configurations (circles).  The dotted line represents the asymptotic
approximation of order 0.
We also plot results obtained by means of the  numerical integration of
Eq.\ (\ref{t1Nb}) when the first-order (dotted line) and second-order (solid
line) asymptotic approximations for $\langle S_N(t) \rangle$ are used. }
\label{fig:t1N2D} 
\end{figure}

\begin{figure}
\begin{center}
\leavevmode
\epsfxsize = 6.5cm
\epsffile{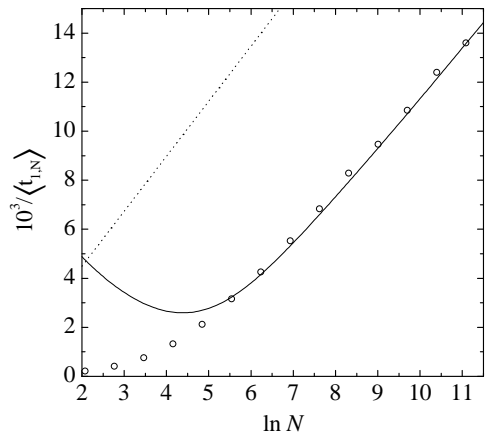}
\end{center}
\caption{ The same as Fig.\ \protect\ref{fig:t1N2D} but for the three-dimensional lattice with
$c=4\times 10^{-5}$. The first-order approximation is out of scale. The simulation
results are averaged over $10^6$ configurations.}
\label{fig:t1N3D}
\end{figure}

In Figs.\ \ref{fig:t1N1D}-\ref{fig:t1N3D}, $\langle t_{1,N}
\rangle$ calculated from Eq.\ (\ref{t1Nc}) is compared with numerical simulation
results.
For the two- and three-dimensional lattices we also show the results obtained by means of
the numerical integration of Eq.\ (\ref{t1Nb}) when  the first- and
second-order asymptotic approximations for $\langle S_N(t) \rangle$ [cf. Eq.\ (\ref{SNt})
with $n=1$ and $n=2$, respectively] are used for $t\ge t_\times\equiv (4/D)\ln N$.
For $t\le t_\times$, the expression  $\langle S_N(t) \rangle=v_0 t^d$
corresponding to the short-time regime is used. For $d=1$, the first- and second-order
results are analytical (see Sec.\ \ref{sec:Rig1D}).
Figures  \ref{fig:t1N1D}-\ref{fig:t1N3D} illustrate the
great  importance of the asymptotic corrective terms in the order-statistics
quantities. The way in which the lines corresponding to  the zeroth-order
approximation run almost parallel to the simulation results indicates that the
corrective term goes essentially as $(\ln N)^{-1}$.
This is confirmed in Sec.\ \ref{sec:Rig1D} where it is found that the rigorous
asymptotic expression for $\langle t^m_{j,N} \rangle$ for the one-dimensional lattice
exhibits corrective terms that decay logarithmically with $N$.

From Eq.\ (\ref{tjNb}) and approximating the difference operator $\Delta^j$ by
the derivative of order $j$,  one finds
\begin{equation}
\langle t_{j+1,N}^m \rangle \simeq
\langle t_{j,N}^m \rangle+
m \frac{\Gamma(1+2m/d)}{\left(\lambda v_0 \right)^{2m/d}(4D)^m}     \;
\frac{(\ln N)^{-1-m}}{j}
\label{tjNc}
\end{equation}
for $j\ll N$, or, in terms of the psi (digamma) function  \cite{abramo},
\begin{equation}
\langle t_{j,N}^m \rangle \simeq
\langle t_{1,N}^m \rangle+
m \frac{\Gamma(1+2m/d)}{\left(\lambda v_0 \right)^{2m/d}(4D)^m}  \;
\frac{\psi(j)-\psi(1)}{(\ln N)^{1+m}} .
\label{tjNcPsi}
\end{equation}
For $1\ll j\ll N$ one gets
\begin{equation}
\langle t_{j,N}^m \rangle \simeq
\langle t_{1,N}^m \rangle+
m \frac{\Gamma(1+2m/d)}{\left(\lambda v_0 \right)^{2m/d}(4D)^m} \;
\frac{\gamma+\ln j}{(\ln N)^{1+m}}
\label{tjNcPsiAsin}
\end{equation}
because $\psi(j)=\ln(j)+ {\cal O}(1/j)$ and $\psi(1)=\gamma$ \cite{abramo}.

Therefore, the
variance $ \sigma^2_{j,N}= \langle t_{j,N}^2 \rangle - \langle t_{j,N} \rangle^2$
is given by
\begin{equation}
 \sigma_{j,N}^2 \simeq
\sigma_{1,N}^2 \simeq
\frac{\Gamma(1+4/d)- \left[\Gamma(1+2/d)\right]^2}{\left(\lambda v_0 \right)^{4/d}
\left(4D \ln N\right)^2 } .
 \label{sigmajN}
\end{equation}
Thus,  the main asymptotic term of the ratio
$\sigma_{j,N}/\langle t_{j,N} \rangle$ is independent of  $j$ and $N$ for large
$N$:
\begin{equation}
 \frac{\sigma_{j,N}}{\langle t_{j,N} \rangle}  \simeq
\frac{\left[\Gamma(1+4/d)- \Gamma^2(1+2/d)\right]^{1/2}}{\Gamma(1+2/d)}
\label{ratiostjN}
\end{equation}

\begin{figure}
\begin{center}
\leavevmode
\epsfxsize = 6.5cm
\epsffile{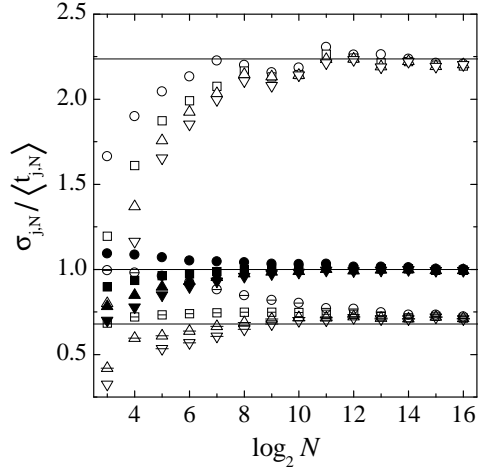}
\end{center}
\caption{ The ratio $\sigma_{j,N}/\langle t_{j,N}\rangle$ , $j=1$ (circles)
$j=2$ (squares), $j=3$ (up triangles)  $j=4$ (down triangles),  $N=2^3,
2^4,\ldots,2^{16}$, for $d=1$ with $c=8\times 10^{-3}$ (hollow
symbols at the top of the figure),  $d=2$ with $c=4\times
10^{-4}$ (filled symbols) and $d=3$ with  $c=4\times 10^{-5}$
(symbols with a bar at the bottom of the figure).
The simulation results are averaged over $10^5$ configurations for $d=1$ and $d=2$
and over  $10^4$ configurations for $d=3$.
The  lines represent the (main order) asymptotic theoretical results, namely,
$\sqrt{5}$ for $d=1$,   $1$ for $d=2$ and $0.678968\cdots$ for $d=3$.}
\label{fig:ratiosigmom}
\end{figure}

In Fig.\ \ref{fig:ratiosigmom} we plot this ratio for the one-, two- and
three-dimensional lattices  for several values of $j$ and  $N$.
The simulation results follow closely the theoretical
predictions.

Finally, note that Eqs.\ (\ref{t1Nc})-(\ref{ratiostjN}) are valid for a given $d$
when  $N\rightarrow \infty$.  
For a given $N$ and  $d   \rightarrow \infty$,  time regimes I and
II shrink,  i.e., $t_{\times}^{\prime}     \rightarrow  0$, so that  $\langle S_N(t) \rangle\sim N
t$ because $\langle S_1(t)\rangle \sim t$ for $d\ge 3$.  
Introducing this relation  into
Eq.\ (\ref{t1Nb}) one gets $\langle t_{1,N}^m \rangle \sim (\lambda N)^{-m}$ for
$d\rightarrow \infty$, as expected.

\section{Order statistics of the one-dimensional trapping process. Rigorous results}
\label{sec:Rig1D}

In this section we obtain the order statistics of the trapping process for the
one-dimensional lattice from the order statistics of the diffusion process in the presence
of two fixed traps.  Let $\overline{t_{j,N}^m}(r)$ be the $m$th moment of the
trapping time of the $j$th particle out of a total of $N$ particles that were initially
placed at distance $r$ from a trap in a given direction and at a distance greater than
$r$ from another trap in the other direction on a line. This quantity is given by
\cite{OS-Yuste} 
\begin{equation} 
\overline{t_{j,N}^m}(r)=  \left(  \frac{r^2}{4D \ln
\kappa N} \right)^{2m} \tau_{j,N}(m), \label{bartmjN}
\end{equation}
where  $ \tau_{j,N}(m)= \tau_{1,N}(m)+ \delta_{j,N}(m)$,
\begin{eqnarray}
\tau_{1,N}(m) =&&
 1+\frac{m}{\ln \kappa  N}
  \left(\frac{1}{2}\ln\ln\kappa  N-\gamma \right)+
    \frac{m}{2\ln^2 \kappa  N}
      \left[1+ \gamma +(1+m)\left(\frac{\pi^2}{6}+\gamma^2\right)         \right.  \nonumber \\
    && \left.
    - \left(\frac{1}{2}+(1+m)\gamma \right)\ln\ln\kappa  N+
		     \frac{1}{4}(1+m) \ln^2\ln\kappa  N \right] +
      {\cal O}\left(\frac{\ln^3\ln\kappa  N}{\ln^3\kappa  N}\right)
				    \; ,
\label{tau1N}
\end{eqnarray}
\begin{equation}
\delta_{j,N}(m)= \frac{m}{\ln \kappa N}\;
 \sum_{n=1}^{j-1} \frac{\delta_n(m)}{n}  ,
\label{deltamjN}
\end{equation}
\begin{eqnarray}
 \delta_n(m)=&&
1+
\frac{m+1}{\ln\kappa N}
\left[(-1)^n\frac{S_n(2)}{(n-1)!}+\frac{1}{2} \ln\ln(\kappa N)-\frac{1}{2(m+1)}-
\gamma \right]
 + {\cal O} \left( \frac{\ln^2\ln\kappa N}{\ln^{2}\kappa N}\right)
\label{Delta}
\end{eqnarray}
and $\kappa=1/\sqrt{\pi}$.

In order to get $\langle t_{j,N}^m\rangle$,  $\overline{t_{j,N}^m}(r) $ is averaged
over the different  positions on which the $N$ particles can be initially placed in an
interval free of traps of size $L$:
\begin{equation}
\overline{\overline{t_{j,N}^m}}(L)=
\frac{2}{L} \int_0^{L/2} dr\; \overline{t_{j,N}^m}(r)=
\frac{1}{2m+1}
\left(\frac{1}{4 D \ln \kappa N}\right)^m
\left(\frac{L}{2}\right)^{2m} \tau_{j,N}(m)   .
 \label{tildetmjN}
\end{equation}
Next,  this quantity is averaged over the size distribution $\eta(L)=\lambda^2 L
\exp[-\lambda L]$ of the intervals that are free of traps \cite[p. 217]{Weiss} to get
the final result
\begin{equation}
\langle t_{j,N}^m \rangle   =
\int_0^{\infty} dL \; \eta(L) \;  \overline{\overline{t_{j,N}^m}}(L)=
 \frac{\Gamma(1+2m)}{(2\lambda)^{2m}}
 \frac{\tau_{j,N}(m)}{\left( 4D\ln \kappa N  \right)^m}
\label{tmjN1Dasin}
\end{equation}
for large $N$ and $d=1$.
In Fig.\ \ref{fig:t1N1D}, the theoretical results given by Eq.\
(\ref{tmjN1Dasin}) are compared with  simulation data.
A behaviour very
close to that found  for traps arranged over a (hyper) spherical surface
\cite{OS-Yuste} is found: the asymptotic corrective terms are not at all negligible even
for very large values of $N$, and the second-order asymptotic expression is an excellent
approximation even for not too large values of  $N$ (say, for $N \gtrsim 100$).

Notice that the approximate result obtained in Eq.\ (\ref{t1Nc}) agrees, for the one-dimensional case, with the main term of Eq.\ (\ref{tmjN1Dasin}). 
This prompts us to investigate to what extent the approximate
procedure of Sec.\ \ref{sec:Rig1D} is able to reproduce the results of the rigorous asymptotic approach. 
The answer is that the two  approaches lead to the same main term (as
we have just discovered) and  to  almost the same first corrective term.
For example, using Eq. (\ref{SNt}) up to first-order corrective terms, one gets for $j=1$ and $m=1$ that
\begin{equation}
\langle t_{1,N}\rangle =
 \frac{1}{2\lambda^{2}  4D\ln \kappa N }
  \left( 1+\frac{\ln \pi -2 \gamma+\alpha+\ln\ln N}{4\ln N}+
         \ldots  \right)
 \label{tmjN1DasinRos}
\end{equation}
 with $\alpha=0$. This expression differs from the rigorous
asymptotic formula (\ref{tmjN1Dasin}) in the value of $\alpha$ only: the exact value is
$\alpha=\ln 2$. Finally,  from Eq.\  (\ref{tmjN1Dasin}) one can also obtain
for $\langle t_{j+1,N}^m\rangle- \langle t_{j,N}^m\rangle $  the  formula
(\ref{tjNc})  which was
obtained in Sec.\ \ref{sec:OSRA} for $d$-dimensional  media.

Finally, from Eq.\ (\ref{tmjN1Dasin}) one gets the variance:
\begin{equation}
\sigma^2_{j,N} =
 \frac{\Gamma(5)\tau_{j,N}(2)-\Gamma^2(3) \tau^2_{j,N}(1)}
 {(2\lambda)^4  (4D\ln \kappa N)^2 }
 \label{sigmajNRig}
\end{equation}
whose main-order asymptotic term reproduces Eq.\ (\ref{sigmajN}) when $d=1$.

\section{Conclusions}
\label{sec:Conclu}
The problem addressed  in this paper is easy to formulate: When a set of $N\gg 1$ diffusing
particles are placed on a site of a $d$-dimensional Euclidean lattice ocuppied by a
random distribution of static traps, how long is the survival time
$t_{j,N}$ of the $j$th trapped particle?
The answer to this order-statistics problem  is given in Eq.\ (\ref{tjNa}) in
terms of the probability $\Phi_{j,N}(t)$  that $j$ particles have been trapped  and
$N-j$ survive by time $t$, which, in turn, can be expressed [cf. Eq.\ (\ref{PhijN})]
exactly in terms of the survival probabily $\Phi_M=\Phi_{0,M}$ that no particle of an
initial set of $M$ ($M=N,N-1,\ldots, N-j$) has been trapped by time $t$.

For the evaluation of $\Phi_N(t)$ we resorted to the Rosenstock approximation generalized to the case of $N\gg 1$ particles.
This approximation is good for small concentrations of traps and small times.
Its  range of applicabilty  depends logarithmically on
$N$,  improving slightly for $d=1$ and worsening slightly for $d=3$ when $N$ increases.
Analytical expressions for the  main asymptotic term of
$m$th moment of $t_{j,N}$ and its variance $\sigma^2_{j,N}$ for $d$-dimensional
Euclidean media have been found by assuming that the density of traps is such that  the
contribution of $\Phi_N(t)$ to  $ \langle t^m_{j,N} \rangle$ is negligible in the time
regimes I and III. It was found that  $ \langle t^m_{1,N} \rangle \sim \left(
\lambda^{2/d} \ln N \right)^{-m} $  and that the ratio $\sigma_{j,N}/\langle t_{j,N}
\rangle$  is not at all  negligible.  In fact $\sigma_{j,N}$ is larger than the
difference  $\langle t_{j+1,N} \rangle- \langle t_{j,N}\rangle $, which implies that it
is not possible to infer with certainty the order $j$ of a trapped particle from the
time at which it is trapped.  However, this ratio  discriminates clearly  the
dimension of the Euclidean media in which the particles diffuse. 
This leads us to consider the possibility that  this ratio could serve to estimate the  dimension of fractal (disordered) media in a dynamical way.

For the one-dimensional lattice,  the
previous solution of the order-statistic diffusive problem for a given configuration (no randomly distributed) of traps  has been used to obtain second-order asymptotic rigourous
expressions for $\langle t^m_{j,N}\rangle$ and the variance $\sigma^2_{j,N}$.
For $d\ge 2$ we resorted to numerical integration  to obtain  higher-order
estimates. This numerical procedure leads to excellent results, but it is limited
to not too large values of $N$ and $j$ because otherwise the binomial term that appears in
Eq.\ (\ref{PhijN}) [or in Eq.\ (\ref{tjNb})] becomes intractably large.  In all the
cases studied, there became clear the great importance of the corrective terms in the
asymptotic expressions of the moments of the  order-statistics quantities since the
$m$th corrective term decay mildly as roughly the $m$th power of the logarithm of $N$.
This characteristic behaviour is shared with other cases with different configurations
of traps (e.g., fixed traps)  and  substrates (e.g., fractal media).

We shall finish by mentioning some open problems.  First, it would very interesting to
estimate the time $t_{N,N}$ by which all the particles are eventually absorbed. Notice
that the formulae of Secs.\ \ref{sec:OSRA} and \ref{sec:Rig1D} are not suitable for this
purpose as they are valid for estimating $t_{j,N}$ when $j\ll N$ only. Also, it would be
interesting to describe the order statistic of the trapping problem for a trap
concentration small enough for the trapping process to take place  mainly inside the
Donsker-Varandhan time regime. The recent analysis of Barkema et. al. \cite{DV-BBB} on
the crossover from the Rosenstock behaviour to the Donsker-Varandhan behaviour should
facilitate this task. Finally, it would be desirable to extend the results of the
present paper to fractal substrates. To this end, the recent results obtained in Ref.
\cite{SNtFrac} on the territory explored by a set of random walkers in fractal media
should be very useful.

\acknowledgments

This work has been supported by the Ministerio de Ciencia y Tecnolog\'{\i}a (Spain) through
Grant No. BFM2001-0718.
SBY is also grateful to the DGES (Spain) for a sabbatical grant (No.\ PR2000-0116) and to  Prof.
K. Lindenberg and the Department of Chemistry and Biochemistry of the  University of
California San Diego for their hospitality.



\begin{references}
\bibitem {nota1} Notice that in the value of  $\tilde p$ given in Ref.
\protect\cite{SNtEucRegular} there were two mistakes:  a 2 should be changed to a 6 and the
symbol $t$ should be deleted. This symbol $t$ should also be deleted from the 
expression for $\tilde p$ given in \protect\cite{SNtEucRapid}.
\end{references}
\end{document}